\documentclass{aa}  
\usepackage{graphicx,psfig}
\usepackage{txfonts}
\newcommand{\Msun} {M$_\odot$}
\newcommand{\Lsun} {L$_\odot$}
\newcommand{\Rsun} {R$_\odot$}

\newcommand{\Lstar} {L$_\star$}
\newcommand{\Mstar} {M$_\star$}

\newcommand{\Myr} {M$_\odot$/yr}

\newcommand{\Ha} {H$_\alpha$}

\newcommand{\Brg} {Br$_\gamma$}
\newcommand{\Lbg} {L($Br_\gamma$)}

\newcommand{\Macc} {$\dot M_{acc}$}
\newcommand{\Lacc} {L$_{acc}$}
\newcommand{\Roph} {Ophiuchus}
\newcommand{\Wcirc} {W(Br$_\gamma$)$_{circ}$}
\newcommand{\Wphot} {W(Br$_\gamma$)$_{phot}$}
\newcommand{\Wobs} {W(Br$_\gamma$)$_{obs}$}

\newcommand{\simless}{\mathbin{\lower 3pt\hbox
      {$\rlap{\raise 5pt\hbox{$\char'074$}}\mathchar"7218$}}} %< or of order
\newcommand{\simgreat}{\mathbin{\lower 3pt\hbox
     {$\rlap{\raise 5pt\hbox{$\char'076$}}\mathchar"7218$}}} %> or of order

\begin{document}
\title{Accretion Rates in Herbig Ae stars
%\thanks{Based on observations collected at the Paranal Observatory, European Southern Observatory, Chile, as part of ESO programme 073.C--0184}
}
%   \subtitle{I. Overviewing the $\kappa$-mechanism}

   \author{R. Garcia Lopez \inst{1,2},
          A. Natta \inst{1}, L. Testi \inst{1} 
          \and
          E. Habart \inst{1,3}
          }

   \offprints{A. Natta, natta@arcetri.astro.it}

   \institute{INAF - Osservatorio Astrofisico di Arcetri,  
              Largo Fermi 5, 50125 Firenze, Italy\\
%              \email{wuchterl@amok.ast.univie.ac.at}
         \and
%INAF - Osservatorio Astronomico di Roma,
%via Frascati 33, 00040 Monte Porzio, Italy \\
 Universit\`a degli Studi di Roma "Tor Vergata" - Dipartimento di Fisica, via della Ricerca Scientifica 1, 00133 Roma, Italy\\
%             \email{c.ptolemy@hipparch.uheaven.space}
          \and
Universit\'e Paris-Sud, 91405 Orsay, France
%   \email{ emilie.habart@ias.u-psud.fr
             }

   \date{Received ... ; accepted ...}

\abstract
{
{Accretion rates from disks around pre-main sequence stars are of importance
for our understanding of planetary formation and disk evolution. We
provide in this paper estimates of
the mass accretion rates in
the disks around a large sample of Herbig Ae stars.}
{We obtained medium resolution
2 $\mu$m spectra  and
used the results to compute
values of   \Macc\  from the 
measured luminosity of the \Brg\ emission line, using
a well established   correlation between L(\Brg) and the accretion luminosity
\Lacc.} 
{We find that 80\% of the stars, all of which have evidence of an associated
circumstellar disk, are accreting matter, with rates 
$3\times 10^{-9}\simless \dot M_{acc} \simless 10^{-6}$ \Msun/yr; for 7
objects, 6 of which are located on the ZAMS in the
HR diagram, we do not detect any line emission. Few HAe stars
(25\%) have \Macc$>10^{-7}$ \Msun/yr. }
{In most HAe stars the accretion rate is sufficiently low that the gas in
the inner disk, inside the dust evaporation radius, is
optically thin and does not prevent the
formation of a  puffed-up rim, where dust is directly exposed to the
stellar radiation.
When compared to the \Macc\ values found for lower-mass stars in the star
forming regions Taurus and Ophiuchus, HAe stars have on average higher
accretion rates than solar-mass stars; however, there is a lack of very strong accretors
among them, probably due to the fact that they are on average
older.}
}
%5 {} token are mandatory
%\keywords{Herbig Ae stars, Accretion, Circumstellar disks}
\maketitle
%
%________________________________________________________________

\section{Introduction}

Pre-main sequence stars accrete matter from their circumstellar disks
for a large fraction of their life, until the  disk is accreted
entirely
onto the star or is dissipated by a number of potentially competing
processes, such as dynamical perturbations by close companions, photoevaporation by the host or nearby  stars, or planet formation. 
Although the amount of matter accreted is small compared to the mass of the central object,
accretion of matter through the disk has important consequences on the physics of the
disk itself and on the properties of the planetary systems that the disk
may form.

The mass accretion rate through the disk
is reasonably well known for pre-mains sequence
stars with masses from about one solar mass (T Tauri stars or TTS), 
down to brown dwarfs (BDs). Results can be found, e.g., in Gullbring
et al.~\cite{Gea98}, Muzerolle et al.~(\cite{Mea98}, \cite{Mea03},
\cite{Mea05}), White and Ghez (\cite{WG01}), Natta et al.~(\cite{Nea04},
\cite{Nea06}), Mohanty et al.~(\cite{Subu05}).

%They may lay in the physical processes that control the momentum and mass
%transfer in the disk. It has been proposed, for example,   (ref) that 
%higher level of
%X-ray radiation from the star may increase the disk ionization, 
%making the magnetorotational instability mechanism (Balbus \& Howley) more
%efficient. Alexander \& Armitage (\cite{AA06}) have shown that the steep
%correlation \Macc-\Mstar\   can be accounted for if the viscous time scale
%at the time of the disk formation scales as 1/\Mstar. Dullemond, Natta \& Testi
%(\cite{Dea06}) have shown how the correlation results naturally in a core collapse model for the star formation if  cores of different mass 
%rotate at the same fraction of their break-up speed.
%Padoan et al. (\cite{Pea04}) propose that  continuous Bondi-Hoyle
%accretion from the surrounding environment controls the disk accretion rate. 

%Stars of intermediate mass may provide important clues to this debate, as they
%may increase the range of luminosity over which \Macc\ is detrmined by very large
%factors.  Also, the sample of Herbig Ae  (HAe) stars with associated disks,
%which are the more massive analogous of the classical TTS, cover a 
%wide range of environments and ages. Their lack of homeneity, when compared to
%the samples of lower mass objects studied so far, may turn out to be advantageous,
%rather than disadvantageous.

It has been much more difficult to  determine
\Macc\  in stars more massive than about 1 \Msun.
Recently
Calvet et al.~(\cite{Cea04}) obtained  the accretion rate in nine young
pre-main sequence stars
with masses in the interval 1.5--3.7 \Msun\ (intermediate mass TTS or IMTTS)
using HST and optical spectra. They found an average accretion rate
of $\sim 3\times 10^{-8}$ \Myr, with a spread of an order of magnitude at least.
%The correlation \Macc--\Mstar shown by TTS and BDs
%seem to  extend with the same slope to  higher masses.
Muzerolle et al.~(\cite{Mea04}) derived a similar value of \Macc\ for the $\sim 2$
\Msun\ star UX Ori by fitting  its \Ha\ profile with magnetospheric accretion models.

We report in this paper measurements of the accretion rate
in a large sample of Herbig AeBe objects (intermediate mass pre-main sequence
stars),
mostly of spectral type  A (HAe).
We  derive the accretion rate  from the luminosity
of the \Brg\ line seen in emission in pre-main sequence accreting objects.
Calvet et al.~(\cite{Cea04}) have shown that 
the correlation between the hydrogen IR line luminosity and the accretion luminosity
found in  TTS  and BDs (Muzerolle et al.~\cite{Mea98};
Natta et al.~\cite{Nea04}) extends to their sample
of IMTTS (see also van den Ancker ~\cite{vda04}). 
Establishing 
the luminosity of the hydrogen near-IR recombination lines as proxy of
the accretion luminosity has been extremely important, since it has made possible
to derive  accretion rates with reasonably good accuracy
in objects where measuring the UV veiling is challenging. This is the case, for
example, of  very young star-forming regions, such as \Roph, where
the high extinction is the limiting factor
(e.g., Natta et al.~\cite{Nea06}), but also of the Herbig AeBe stars,
where it is difficult to separate the UV excess emission from the
hot photospheric continuum.

%This sample, however, is small and restricted to objects in a couple of
%star-forming regions.
The main purpose of this work  is to estimate \Macc\ 
in  objects with disks for which we are currently obtaining
near-IR interferometric data with the ESO VLTI (AMBER GTO\footnote{
http://www.eso.org/observing/proposals/gto/amber/}; see also
Malbet et al.~\cite{Malb06}, Tatulli et al.~\cite{Tea06}).
Knowledge of \Macc\ is crucial
for  understanding  the disk geometry, in particular   the nature of
the disk inner rim (e.g., Natta et al.~\cite{Nea01}; Dullemond
et al.~\cite{DDN01}; Isella \& Natta \cite{IA05}), which
can only form if the accretion rate is lower than
a critical value of the order of
 $10^{-7}$ \Msun/yr  (Muzerolle et al.~\cite{Mea04}).
For low \Macc, the inner gaseous disk is optically thin and
does not prevent the direct stellar irradiation of the inner dust edge of
the disk, thus giving raise to the puffed-up rim. 
Secondly, we  are interested in  comparing the results to the accretion properties
of the lower-mass objects studied in Taurus and Ophiuchus.
In both regions it has been found 
that the mass accretion rate \Macc\ is a strong function
of the mass of the central object, increasing 
roughly as \Mstar$^{1.8-2.0}$ (e.g., Natta et al.~\cite{Nea06}
and references therein; but see also Clarke \& Pringle \cite{CP06}).  
The causes of this behaviour are not understood, and are matter
of debate (Muzerolle et al.~\cite{Mea05}; Natta et al.~\cite{Nea06};
Alexander \& Armitage \cite{AA06}; Dullemond et al.~\cite{Dea06}).
The HAe stars are scattered in different  regions of the sky
and are, on average, older than objects in Ophiuchus and Taurus;
they may provide interesting information on the dependence of the
disk accretion on age and environment.

The outline of this paper is as follows. The observations and data reduction are described in \S2; method and  results are presented in \S3 and discussed in \S4.
Conclusions follow in \S 5.

\section {Observations and data reduction}

Medium resolution (R$\sim$9000) spectra of all stars in our sample were obtained with
the ISAAC spectrograph at the Antu 8.2m VLT unit telescope in service mode in spring 2004. 
The spectral region  covered by our spectra included the Br$\gamma$ and the H$_2$ (1--0)S(1) lines, 
which was not detected in any of the objects of
the sample. All spectra were obtained in moderately poor conditions (seeing $\sim$1.5 arcsec and non photometric transparency), 
as these were adequate for the scope of the project. 
As a consequence, we did not attempt to obtain an accurate flux calibration of the spectra.
%The total on-source integration time per target was approximately 20min. {check}

The raw data were wavelength calibrated using daytime arcs and rectified using data from the ISAAC calibration plan. After aligning and coadding the dithered integrations on each target, the spectra were extracted and telluric absorption and
instrumental throughput were corrected using observations of telluric standards obtained by the observatory staff during the same nights as the science observations. 
The details of the ISAAC calibration plan, the reduction  process and the instrument characteristics can be found on the ESO web pages\footnote{http://www.eso.org/instruments/isaac}.

The portion of the spectra around  \Brg\ are shown and briefly discussed in the appendix. In each spectrum we measured the equivalent width of the Br$\gamma$ line; the measurements are given in Tables~1 and~2, together with the 1$\sigma$ uncertainties.

Tables~1 and 2 report also the star name and properties. Spectral type and distance are compiled from the literature
(e.g., van den Ancker et al.~\cite{vda97}, \cite{vda98}; Testi et al.~\cite{Tea98}; 
Meeus et al.~\cite{Meeus01}; Rodgers \cite{Rod01};
Hernandez et al.~\cite{Hea05},  and references therein).
Extinction and luminosity are computed from the V, (B-V) magnitudes as described in
Testi et al.~(\cite{Tea98}); stellar masses are derived by comparing the location of each star on the HR diagram to the evolutionary tracks of Palla and Stahler (\cite{PS93}).

\section {Method and results}

We  derive the accretion luminosity \Lacc\ of each star from
the luminosity of \Brg, following the relation derived by
Calvet et al.~(\cite{Cea04}) for a sample of objects ranging in mass from
about 4 \Msun\ to BDs:
\begin{equation}
\log L_{acc}/L_\odot = 0.9 \times \large(\log (L(Br_\gamma)/L_\odot) +4\large) -0.7
\end{equation}
where \Lbg\ is the luminosity of the \Brg\ line emitted by the circumstellar
gas.
Note that Eq.(1) is an empirical relation, where
\Lacc\ is measured from the observed veiling, i.e., independently of any
assumption on the  origin of the \Brg\ emission.
It has been derived from objects with
L(\Brg) between $\sim 3\times 10^{-6}$ and
$\sim 3\times 10^{-3}$ \Lsun; the
scatter of \Lacc\ for any given L(\Brg)
is typically a factor $\pm$3 (see also van den Ancker \cite{vda04}).
%The validity of Eq.(1) for  objects with \Brg\ luminosity  
%$\simless 10^{-3}$ \Lsun\ has been confirmed by van den Ancker (\cite{vda04}).

In order to compute \Lbg, we need to extract the equivalent width of
the circumstellar component of the line from the observed one. In HAe
stars, the observed flux is
the sum of the stellar photospheric emission, which
for A stars shows broad and rather strong absorption in the hydrogen
lines, of the emission of the circumstellar gas and of the emission
of the disk. Following Rodgers (\cite{Rod01}), we  assume that 
the star dominates the observed V emission. 
The equivalent width of the photospheric component, corrected for the
K-band veiling, is computed for each object with spectral type earlier
than A8 using the template spectra of Rodgers (\cite{Rod01}). 
For
objects of later spectral type, for which we did not have
spectra of template stars of suitable resolution,  we compute
\Wcirc\ from the expression:
%the flux of the circumstellar \Brg, integrated over the line profile, as:
%
%\begin{equation}
%F(Br_\gamma)= \int{\big[1-{{F_\star^K \phi(\lambda) + 
%F_{disk}^K}\over{F_\star^K + F_{disk}^K}}\big] d\lambda} \>\> -(F_\star^K + F_{disk}^K)\> EW_{obs}(Br_\gamma)
%\end  {equation} 
%where $F_\star^K$ is the continuum photospheric flux computed from the
%extinction-corrected V magnitude and the photospheric color index (V-K)$_0$
%appropriate for the star spectral type; $\phi(\lambda)$ is the \Brg\
%photospheric profile, derived from a spectroscopic template (see Fig.~\ref{templates}); $F_{disk}^K$ is the disk continuum emission, computed subtracting
%from the observed K magnitude  the photospheric one.
%We can then write 
%the equivalent width of the circumstellar \Brg\ component as:
\begin{equation}
W(Br_\gamma)_{circ}= W(Br_\gamma)_{obs} -W(Br_\gamma)_{phot}\> 10^{-0.4 \Delta K}
\end{equation}
where \Wobs\ is the observed equivalent width, 
\Wphot\ the equivalent width of the photospheric \Brg\ line
of a template with the  same spectral type (see Table~\ref{table_templates};
values from Rodgers \cite{Rod01} and Wallace \& Hinkle \cite{WH97}) and $\Delta K$ the disk continuum emission, computed subtracting
from the observed K magnitude  the photospheric one.
All the relevant parameters used in the derivation are given in Table~\ref{table_hae}.
 K magnitudes are from 2MASS \footnote{This publication makes use of data
 products from the Two Micron All
Sky
Survey, which is a joint project of the University of Massachusetts and
the Infrared Processing and Analysis Center/California Institute of
Technology, funded by the National Aeronautics and Space Administration
and the National Science Foundation.
}; note that none of the quantities used in eq.(2)
has been measured simultaneously.

Of the 32 targets, 4 are very luminous objects of spectral type early B.
Three of them have \Brg\ in emission, one in absorption.
 Table~\ref{table_hbe} refers in Column 3 a lower limit to
the line  luminosity, 
derived from the observed \Brg\ equivalent width, i.e., without
correcting for the photospheric absorption.
For the three stars where it is seen in emission L(\Brg)$\simgreat 0.1-0.2$ \Lsun.
The validity of Eq.(1) for these very luminous objects is very doubtful,
and we will not consider them further in the following of this paper.

\begin{figure}[h!]
\begin{center}
\leavevmode
\end{center}
%\centerline{ \psfig{file=ps.lacc_hae,width=9cm,angle=0} }
\centerline{ \psfig{file=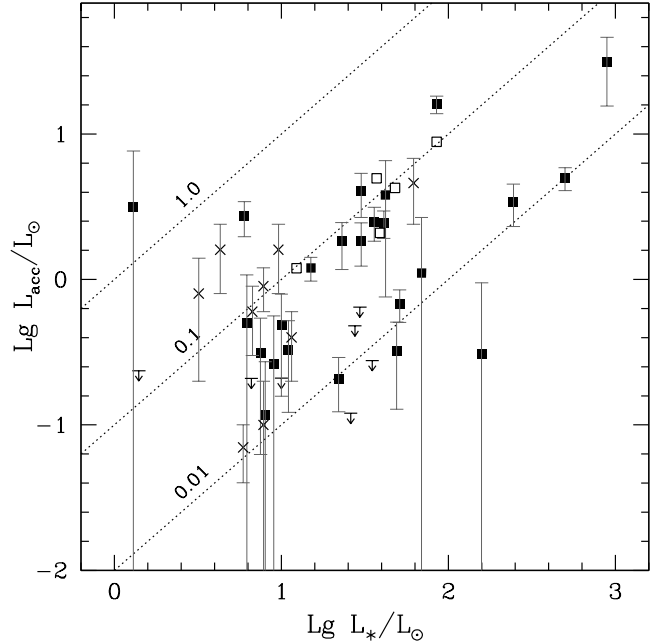,width=9cm,angle=0} }
\caption{Accretion luminosity as a function of \Lstar\ for our sample (filled squares) and the
Rodgers (\cite{Rod01}) objects (open squares); arrows show upper limits. For comparison, we show also the IMTTS \Lacc\ measured by Calvet et al.~\cite{Cea04} from veiling (crosses).}
\label{fig_lacc}
\end{figure}

After  subtraction of the photospheric component, 24 of the 28 HAe
stars have circumstellar \Brg\  emission; for these,
we compute \Lbg\ 
from \Wcirc, the continuum flux derived from the extinction-corrected
K magnitude and the distance of the star.
\Lacc\ is then obtained from Eq.(1).
For the remaining 4, where no circumstellar  emission is detected,
we estimate that 
\Wcirc $\simless$ 2 \AA, and  compute upper limits to \Lacc\ accordingly.
For HD~98922 and HD~144432, \Lacc\ refers to
the quoted distance lower limit, and is therefore also a lower limit to the true value.
The results are given in Table~\ref{table_hae} and shown in Fig.~\ref{fig_lacc} as a function of the stellar luminosity.
Note that Eq.(1) has been derived for objects with L(\Brg)$\simless 3\times 10^{-3}$ \Lsun; 2 objects (HD~98922 and VV Ser) have 
larger L(\Brg),  and the use of Eq.(1) should be viewed with some caution.
 
The mass accretion rate  follows
from the relation $\dot M_{acc}=L_{acc} \, R_\star/(GM_\star)$, and is
shown in Table~\ref{table_hae}, Column (13).

The uncertainties on \Lacc\ and \Macc\ are dominated in most cases by the uncertainties 
deriving from the subtraction of the veiled
photospheric absorption component from the observed spectrum. The error bars shown
in Fig.~\ref{fig_lacc} correspond to an uncertainty on the resulting
\Wcirc\ of  $\sim\pm$2 \AA, which we consider realistic in the majority of
cases.
%The errors deriving from the uncertainties on the correlation between \Lbrg\ and \Lacc\ and on the ratio \Mstar/Rstar\ are generally smaller,
% with the exception of those
%few stars with uncertain distance or spectral type.

For any given object, variability of lines and continuum can also be large.
There are no systematic studies of time variability of the \Brg\ equivalent
width in our sample stars, but one can compare our measurements with
the Rodgers (\cite{Rod01}) results from lower resolution
spectra (R$\sim 500-2000$) for the 7 stars we have in common. The differences
are within the errors for 3 of them, but larger than 2 \AA\ in 2, one
being R CrA. 
This confirms the well known fact that the accretion rate in pre-main
sequence stars varies and that one can only estimate  
the ``typical" value  of any individual object if the results of 
long term time monitoring are available. However, if one is interested in the  properties 
of a large sample of stars, as in this work, one can safely assume that
their time variability will  increase the spread of the determinations, 
but not bias the average values.

\begin{figure}[ht!]
\begin{center}
\leavevmode
\end{center}
%\centerline{ \psfig{file=ps.macc_hist,width=9cm,angle=0} }
\centerline{ \psfig{file=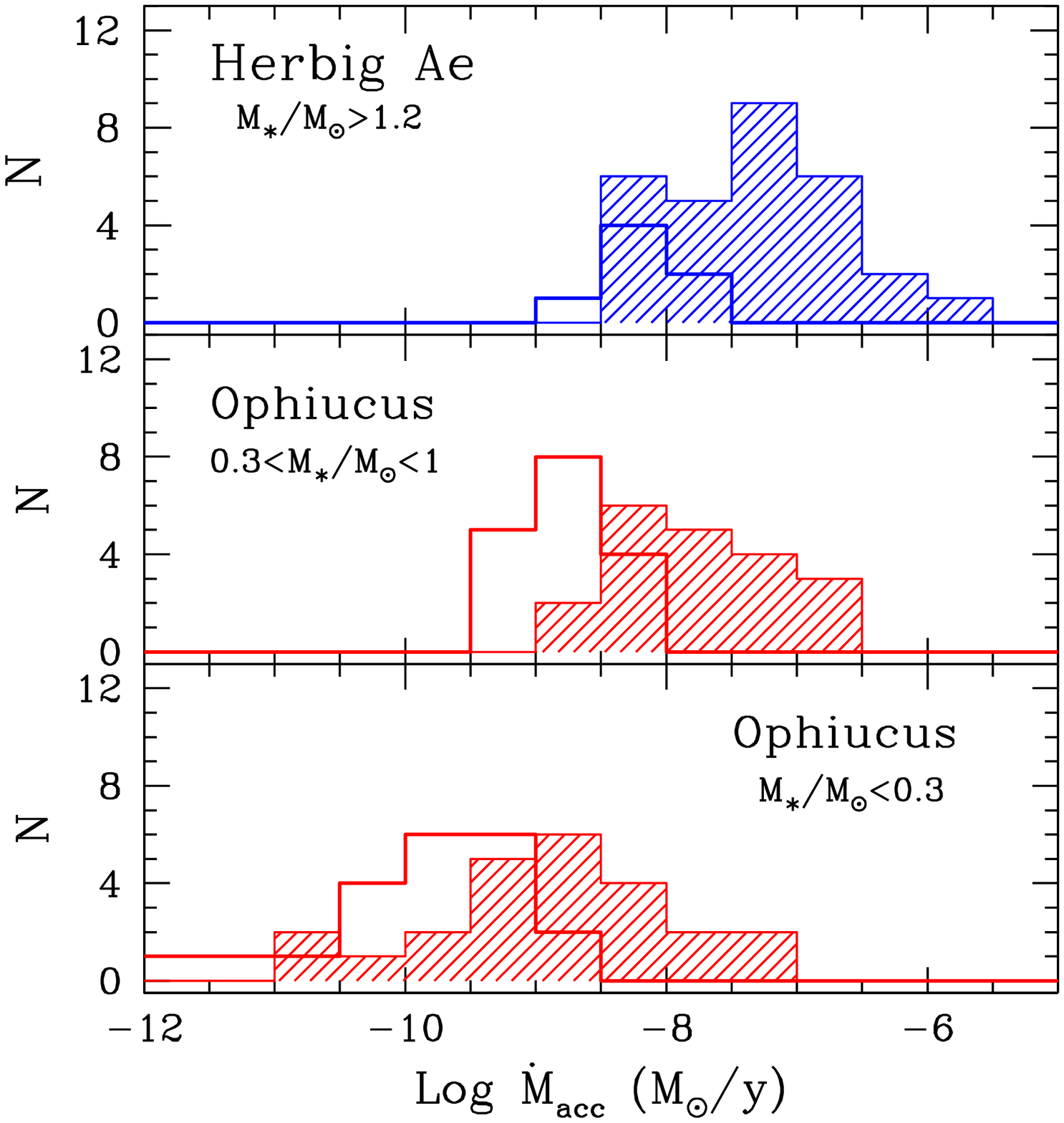,width=9cm,angle=0} }
\caption{Distribution of the mass accretion rate in HAe stars
(top Panel). The shadowed boxes show actual measurements, the empty
boxes (thick histogram) the distribution of the upper limits.
The mid and bottom panels show the distribution of \Macc\ for Class II 
objects
in Ophiuchus with mass $0.3 <M_\ast/M_\odot <1$ and $M_\ast/M_\odot \le 0.3$,
respectively (data from Natta et al.~\cite{Nea06}).
}
\label{fig_hist}
\end{figure}

\begin{figure}[ht!]
\begin{center}
\leavevmode
%\centerline{ \psfig{file=ps.macc_all,width=9cm,angle=0} }
\centerline{ \psfig{file=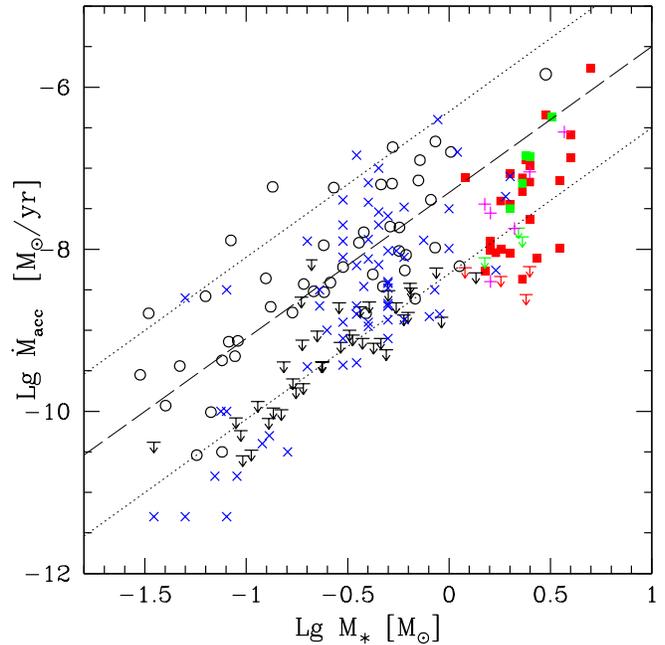,width=9cm,angle=0} }
\end{center}
\caption{Mass accretion rate as a function of \Mstar\ for  HAe stars
(filled squares and upper limits), Ophiuchus Class II objects (open dots and 
upper limits; Natta et al.~\cite{Nea06}), 
Taurus CTTS and BDs (crosses; Muzerolle et al.~\cite{Mea05}, 
Mohanty et al.~\cite{Subu05},
Calvet et al.~\cite{Cea04} and references therein), 
IMTTS in Orion (plus signs, Calvet et al.~\cite{Cea04}).}
The dashed line shows the best-fit to the Ophiuchus data
(\Macc $\propto$ \Mstar$^{1.8}$); the dotted lines are $\pm$1 dex.
\label{fig_macc}
\end{figure}

\begin{table*}
\caption{Herbig Ae properties -- Column 1: star name; Column 2: spectral type; Column 3: V magnitude; Column 4: K magnitude; Column 5: V extinction; Column 6: distance; Column 6: effective temperature; Column 8: stellar luminosity; Column 9: stellar mass; Column 10: observed \Brg\ equivalent width; Column 11: equivalent width of the circumstellar \Brg; Column 12: accretion luminosity; Column 13: mass accretion rate.}
\label{table_hae}
\centering
\begin{tabular}{ l c c c c c c c c c c c  c}     % 13 columns
\hline\hline
(1) & (2)   & (3)& (4) & (5) & (6)& (7)      & (8)      & (9)      & (10)& (11)& (12)& (13)\\

 name  & ST &   V & K& A$_{\rm V}$  & D  & T$_{eff}$& \Lstar& \Mstar& Br$_\gamma$(obs)& Br$_\gamma$(circ)& Log L$_{acc}$& Log $\dot M_{acc}$\\
       &    & (mag)& (mag)& (mag)& (pc)& (K)& (\Lsun)& (\Msun)& (\AA)& (\AA)& (\Lsun)& (\Msun/yr)\\
\hline
 \object {HD 149914} & B9 &6.75& 5.69& 0.9  &165 & 10600   & 158      & 3.5 & 6.9$\pm$0.2& -0.8 & -0.51 & -7.99      \\
\object {HD 179218} & B9 & 7.2 & 5.99& 1.8 & 240& 10600   & 500      & 4   & 1.3$\pm$0.2 & -10.1 & 0.70 & -6.59       \\
\object  {HD 176386} & B9& 7.30 & 6.69& 0.5 & 140& 10600   & 49       & 2.7  & 6.2$\pm$0.2& -3.1    &  -0.49& -8.11 \\
\object  {HD 97300}  & B9 & 9.00& 7.15& 1.2 & 190& 10600   & 35       & 2.5      & 9.1$\pm$0.2& -- &$<$-0.5 &$<$-8.18  \\
\object  {HD 98922}  & B9 & 6.77& 4.28& 0.3 & $>$540& 10600& $>$890   & 5: &-2.4$\pm$0.2 & -3.7    &  1.50& -5.76    \\
\object  {VV Ser}    & B9 &11.92 & 6.32& 3.1 &260 & 10600 & 85   & 3.1   &-9.0$\pm$0.2& -13.0 & 1.20& -6.34        \\
\object  {51 Oph}    & B9.5&4.81& 4.29& 0.1 & 131& 10200   & 245      & 4  &1.3$\pm$0.3& -5.6 & 0.54 &-6.87        \\
\object  {HD 141569} & B9.5&7.00& 6.82& 0.  &99  & 10200   & 22    & 2.3   &6.0$\pm$0.1& -4.5  &  -0.68 & -8.37      \\
\object  {NX Pup}    & A0 &9.96 &6.08& 1.4 & 450& 9840    & 42     & 2.5   &-1.9$\pm$0.2& -2.4     &  0.58& -6.97    \\
\object  {HR 4796}   & A0 &5.78 &5.77 & 0.1 & 67 & 9840   & 26     &2.4    &9.4$\pm$0.2& --& $<$-0.9 & $<$-8.53\\ 
\object  {HD 97048}  & A0 &8.46 &5.94& 1.0 & 180& 9840     & 41    & 2.5     &-6.3$\pm$0.15& -8.4   & 0.39 &-7.17     \\
\object  {HD 104237} & A0 &6.59 &4.58& 0.7 & 116& 9840     & 6     & 2.0     &-4.7$\pm$0.2& -6.7   & 0.43 &-7.45     \\
\object  {HD 95881}  & A1 &8.25 &5.73& 0.3 & 120& 9450    & 10     & 2       &-1.6$\pm$0.2& -2.8   & -0.31 & -8.04     \\
\object  {HD 150193} & A1 &8.88 &5.48& 1.5 & 150& 9450     & 30     & 2.3     &-3.5$\pm$0.15& -5.5  & 0.27 & -7.29      \\
\object  {HD 163296} & A1 &6.87 &4.78& 0.3 & 122& 9450    & 36      & 2.3     &-4.7$\pm$0.1& -6.9  &  0.40 & -7.12     \\
\object  {HD 144667} & A1.5&6.63&6.67&0.1&140& 9300     & 51       & 2.5     &5.8$\pm$0.15& -7.2   & -0.17 & -7.63     \\
\object  {TY Cra}    & A2 &9.42 &6.67& 1.3 & 130& 9120   & 8        & 1.8      &4.7$\pm$0.2& --     &$<$-0.6 &$<$-8.31  \\
\object  {WW Vul}    & A3 &10.51& 7.28&1.0 & 550& 8670   & 30     & 2.4     &-3.6$\pm$0.2&-5.4     & 0.60& -6.89   \\
\object  {HD 169142} & A5 &8.15 & 6.41&0.3 & 145& 8260     & 15    & 1.8     &-5.0$\pm$0.1&-9.7    &  0.08 & -7.40    \\
\object  {KK Oph}    & A6 &10.57 &5.79& 0.44 & 160& 8100  & 6.2   & 1.6     &-1.2$\pm$0.15& -1.5    & -0.30 & -7.91   \\
\object  {HD 139614} & A7 &8.25 & 6.75&0.1 & 150& 7900     & 11    & 1.8     &2.2$\pm$0.2& -3.0& -0.48 &-7.99      \\
\object  {HD 142666} & A8 &8.81 & 6.08&0.8 & 116& 7580     & 8        & 1.6     & 1.0$\pm$0.2&-2.4 &-0.50 & -8.00        \\
\object  {HD 100453} & A9 &7.79 &5.6& 0.0 & 111& 7300     & 9        & 1.7     &1.6$\pm$0.2& -1.5   & -0.58& -8.04     \\
\object  {HD 144432} & A9 &8.16 &5.89& 0.2 & $>$200&7300& $>$23      & $>$2      &-2.1$\pm$0.2& -5.1 &0.26 & -7.07       \\
\object  {HD 135344} & F3 &8.61 &5.84& 0.3 & 84 & 6590     & 8      & 1.5        &0.3$\pm$0.2& -1.3   &-0.93 &-8.27    \\
\object  {T CrA }    & F5:&12.04&6.60& 1.9 & 130& 6400     & 1.4      & 1.2     &9.6$\pm$0.2& --   &$<$-0.6 &$<$-8.20     \\
\object  {R CrA}     & F5:&11.50&2.86& 1.4 & 130& 6400     & 1.3      & 1.2     &-1.2$\pm$0.3& -1.2 & 0.50 & -7.12       \\
\vspace{0.3cm}
\object  {HD 142527} & F6&8.33&4.98&0.7 & 200& 6270     & 69       & 3.5     &0.3$\pm$0.2&-1.2  &0.04  & -7.16\\ 
\object {AB Aur} &  A0&7.05 & 4.2 & 0.5& 140&9840 &48& 2.4& -4.4$\pm$0.15& -5.5&  0.63& -6.85\\
\object {HD 34282}& A0&9.85 & 7.68& 0.6& 400&9840 &30& 2.2&  2.7$\pm$0.17& --   & $<$-0.20 &$<$-7.71\\
\object {RR Tau}& A0&11.0   & 7.39& 1.1& 800&9840& 37& 2.5& -2.56$\pm$0.05& -3.5&  0.70&-6.86\\
\object {HD 35187}& A2&7.78 & 5.91& 0.65&150&9120 &27& 2.3&  7.13$\pm$0.27& --   & $<$-0.3&$<$-7.81\\
\object {UX Ori}&A3&10.0 & 7.21& 0.3& 450&8670& 39& 2.3& -2.4$\pm$0.02& -3.9&  0.32& -7.19\\
\object {HK Ori}&  A5&11.66  & 7.34& 1.2 & 450&8260& 12& 2.0& -1.44$\pm$0.08& -2.2&  0.08& -7.50 \\
\object {HR 5999}& A6&7.04  & 4.39& 0.5& 210&8100& 85& 3.2& -1.74$\pm$0.01& -6.5&  0.95& -6.37\\
\object {CQ Tau} & F2& 9.42& 6.17& 1.9& 130&    &6.6& 1.5&  3.17$\pm$0.18&   --& $<$-0.6 & $<$ -8.08\\
\hline
\end{tabular}
\flushleft Note: The last 9 lines are objects from Rodgers (\cite{Rod01}).
\end{table*}

\begin{table}
\label{table_hbe}
\caption{Herbig Be properties -- 
Column 1: star name; Column 2: spectral type; Column 3: observed \Brg\ equivalent width; Column 4: \Brg\ luminosity.}
\centering
\begin{tabular}{ l c c c}     % 3 columns
\hline\hline
(1) & (2)   & (3)& (4)\\

 name  & ST &   Br$_\gamma$(obs)& Log L(\Brg)\\
       &    &  (\AA)& (L$_\odot$)\\
\hline
\object {V921 Sco}& B0& -17.3$\pm$0.25& $>$-0.7\\
\object {MWC 300}& B0-B1& -10.9$\pm$0.25& $>$-0.8\\
\object {MWC 166}& B0 IV& 2.5$\pm$0.2& --\\
\object {MWC 297}& B1.5& -16.0$\pm$0.2& $>$-0.6\\
\hline
\end{tabular}
\end{table}

\begin{table}
\begin{flushleft}
\caption{Br$_\gamma$ equivalent width of template stars -- Column 1: spectral type; Column 2: adopted \Brg\ equivalent widht.}
\label{table_templates}
\centering
\begin{tabular}{ l  c }     % 3 columns
\hline\hline
(1) & (2)   \\

ST &   W(Br$_\gamma$)\\
   &  (\AA)\\
\hline
B8&  11.4\\
A0& 9.7\\
A2& 13.4\\
A8& 12.0\\
F3& 6.6\\
F6& 5.5\\
\hline
\end{tabular}
\end{flushleft}
\end{table}

\section {Discussion}

Fig.~\ref{fig_lacc} plots \Lacc\ as a function of \Lstar\ for the HAe
observed in this paper. We have 
added six HAe stars (listed at the bottom of Table~1)
taken from the Rodgers (\cite{Rod01}) thesis,
for a total of 36 objects.
%, who measured circumstellar \Brg\ equivalent widths with the same
%procedure used in this paper from low resolution spectra. 
Typically, the HAe stars have \Lacc\ roughly between 10\% and 1\% \Lstar.
There  are very few objects with \Lacc$>$0.1 \Lstar,
contrary to what happens for   lower mass objects in Taurus and Ophiuchus
(see Natta et al.~\cite{Nea06}, Calvet et al.~\cite{Cea04}).

Fig.~\ref{fig_hist}, top Panel, shows the \Macc\ distribution of the HAe
stars; the values of \Macc\ cover the interval 
$\sim 10^{-6}$ -- $\simless 3\times 10^{-8}$ \Msun/yr, but 
few stars (25\%) have \Macc $>10^{-7}$ \Msun/yr;
36\% have \Macc$< 10^{-8}$ \Msun/yr.
The median value is \Macc$=3.5\times 10^{-8}$ \Msun/yr.
These results agree with those obtained by 
Muzerolle et al.~(\cite{Mea04}), who found that  intermediate-mass
objects have typical 
\Ha\ profiles and/or Balmer discontinuity
roughly consistent with the predictions
of magnetospheric accretion models 
with accretion rates $\simless 10^{-7}$ \Msun/yr.
They studied in detail only one object, UX Ori, for which they found a value
\Macc$\sim 10^{-8}$ \Msun/yr, comparable with the value ($\sim 6\times 10^{-8}$
\Msun/yr) we derive from the low resolution spectrum of Rodgers (\cite{Rod01}).

Our results support the conclusion of Muzerolle et al.~(\cite{Mea04}) that
in most HAe stars the accretion rate is sufficiently low that the gas in the
inner disk, inside the dust evaporation radius, is optically thin, allowing the
direct irradiation of the inner dust edge (the rim) by the star.
One of the conditions for the formation of a puffed-up rim (Natta et al.~\cite{Nea01})
seems therefore generally verified.

 Fig.~\ref{fig_macc} plots \Macc\ as a function of
\Mstar\  for the HAe stars, together with objects
in \Roph\ (Natta et al.~\cite{Nea06}), Taurus (Muzerolle et al.~\cite{Mea05}
and references therein; Mohanty et al.~\cite{Subu05}) and  IMTTS in Orion
and Taurus (Calvet et al.~\cite{Cea04}).
Fig.~\ref{fig_hist} compares the \Macc\ distribution of the HAe stars
to that of objects in Ophiuchus in two mass ranges, as labelled.
One can see that
the median  \Macc\ for the HAe stars ($\sim 3\times 10^{-8}$ \Msun/yr)
is about a factor ten higher than for Ophiuchus stars with
$0.3 < M_\star < 1$ \Msun, and a factor ~100
larger than in the very low mass  objects with \Mstar$<0.3$ \Msun.
This is roughly consistent with the predictions of the
 \Macc$\propto$ \Mstar$^{1.8}$ relation derived for sub-solar mass
objects.
A close inspection of Fig.~\ref{fig_macc}, however,
shows   that the HAe stars  lack very 
strong accretors, as already noted in discussing Fig.~\ref{fig_lacc}.
If the upper envelope of the distributions observed in Taurus and Ophiuchus
extended to HAe stars, we would expect a much larger fraction
 of objects accreting at rates higher than $10^{-7}$ \Msun/yr.

It is not easy, at this stage, to understand why this happens.
The HAe stars form a
sample which is not complete nor homogeneous.
The objects are located in a variety of star forming regions, and, 
in some cases, are isolated in the sky. We think that
the most likely interpretation
of their behaviour is ``aging'', since 
the  accretion rate in 
viscous disks with constant $\alpha$  parameter
is expected to deacrease with time roughly as 
$t^{-1.5}$ (Hartmann et al.~\cite{Hea98}).
HAe stars are, on
average, older than Ophiuchus objects, which cluster at ages $\sim 0.5-1$ Myr,
respectively (Palla \& Stahler \cite{PS00}).
Their
nominal ages, obtained by comparing their location on the HR diagram
with Palla \& Stahler (\cite{PS99}) evolutionary tracks, range from
$\sim 1$ to $> 10$ Myr; 
several stars are close to the ZAMS, and it is in fact interesting
that 6/7 stars with  non-detected \Brg\ emission
are ZAMS objects, the exception being HD 35187, with an estimated age
$\sim 8$ Myr.
However,  one should not overinterpret these findings,
as ages of intermediate-mass
stars derived from their location on the HR diagram are
very uncertain.
% (see, e.g., ??). 

It is possible that the steep dependence of \Macc\ 
on \Mstar\ does not extend to objects more massive than, 
say, $\sim 1$ \Msun, reflecting a change in the physical processes that 
control  disk evolution
and/or formation, and which have been discussed in a number of
recent papers (e.g., Muzerolle et al.~\cite{Mea03}, Natta et al.~\cite{Nea06},
Padoan et al.~\cite{Pea04}, Alexander \& Armitage \cite{AA06},
Dullemond et al.~\cite{Dea06}).
Unfortunately, given the lack of completeness and the selection biases
that affect it, it is difficult to establish the
relevance of the HAe sample for this ungoing  discussion. 
It is worth to remind, in this context, the
well  known bias due to  the spectral type selection, which
introduces an age-mass correlation in the sample (e.g., van Boekel et al.~\cite{vBea05})
and automatically excludes young, relatively low mass ($\simless$ 2.5 \Msun)
objects.

\section{Conclusions}

We have reported in this paper measurements of the disk 
mass accretion rates in
a sample of 36 Herbig Ae stars. The values of \Macc\
are derived from the measured luminosity of the \Brg\ emission line, using
the  correlation between L(\Brg) and the accretion luminosity
\Lacc, established by Muzerolle et al.~(\cite{Mea98}) and Calvet et al.~(\cite{Cea04}).
Note that it is an empirical  correlation,  which makes no assumptions 
on the  origin of \Brg.

We find that 80\% of the stars, all of which have evidence of an associated
circumstellar disk, are accreting matter, with rates 
$3\times 10^{-9}\simless \dot M_{acc} \simless 10^{-6}$ \Msun/yr; for 7
objects, 6 of which are located on the ZAMS in the
HR diagram, we do not detect any line emission. Few HAe stars
(25\%) have \Macc$>10^{-7}$ \Msun/yr. 
These results support the conclusion of Muzerolle et al.~(\cite{Mea04})
that in most HAe stars the accretion rate is sufficiently low that the gas in
the inner disk, inside the dust evaporation radius, does not prevent the
formation of the  puffed-up rim, proposed by Natta et al.~(\cite{Nea01})
and Dullemond et al.~(\cite{DDN01}).

When compared to the \Macc\ values found for lower-mass stars in the star
forming regions Taurus and Ophiuchus, HAe stars have on average higher
accretion rates; however, there is a lack of very strong accretors
among them,
as  shown by the  very few objects with \Lacc $>$ 10\% \Lstar.
It seems likely that this is because on average HAe stars
are older than TTS in Ophiuchus or Taurus. However, 
ages of intermediate-mass stars derived from the HR diagram location
are very uncertain; this, 
and  the  various selection biases which plague the HAe sample,
make the comparison very difficult.

When we consider individual stars, 
the largest uncertainties on  \Macc\  come from line variability,
as in TTS,
but also from  the
presence of a photospheric absorption component of \Brg\ in early-type
stars, which needs
to be subtracted from the observed one, and from the K-band
excess emission. 
A better characterization of the photospheric profiles of template stars
would be very valuable.

\begin{acknowledgements}
We thank the ESO staff for competent and efficient support during the preparation
of the observing blocks and for the excellent data delivered for this service observing programme.
The spectroscopic observations of template stars were kindly provided by Bernadette Rodgers. 
This publication makes use of data products from the Two Micron All Sky 
Survey, which is a joint project of the University of Massachusetts and 
the Infrared Processing and Analysis Center/California Institute of 
Technology, funded by the National Aeronautics and Space Administration 
and the National Science Foundation.
This project was partially supported by MIUR grant 2004025227/2004.
\end{acknowledgements}

\appendix
\section{\Brg\ Spectra}

\begin{figure*}[ht!]
\begin{center}
\leavevmode
\end{center}
%\centerline{ \psfig{file=ps.spectra_all,width=18cm,angle=-90} }
\centerline{ \psfig{file=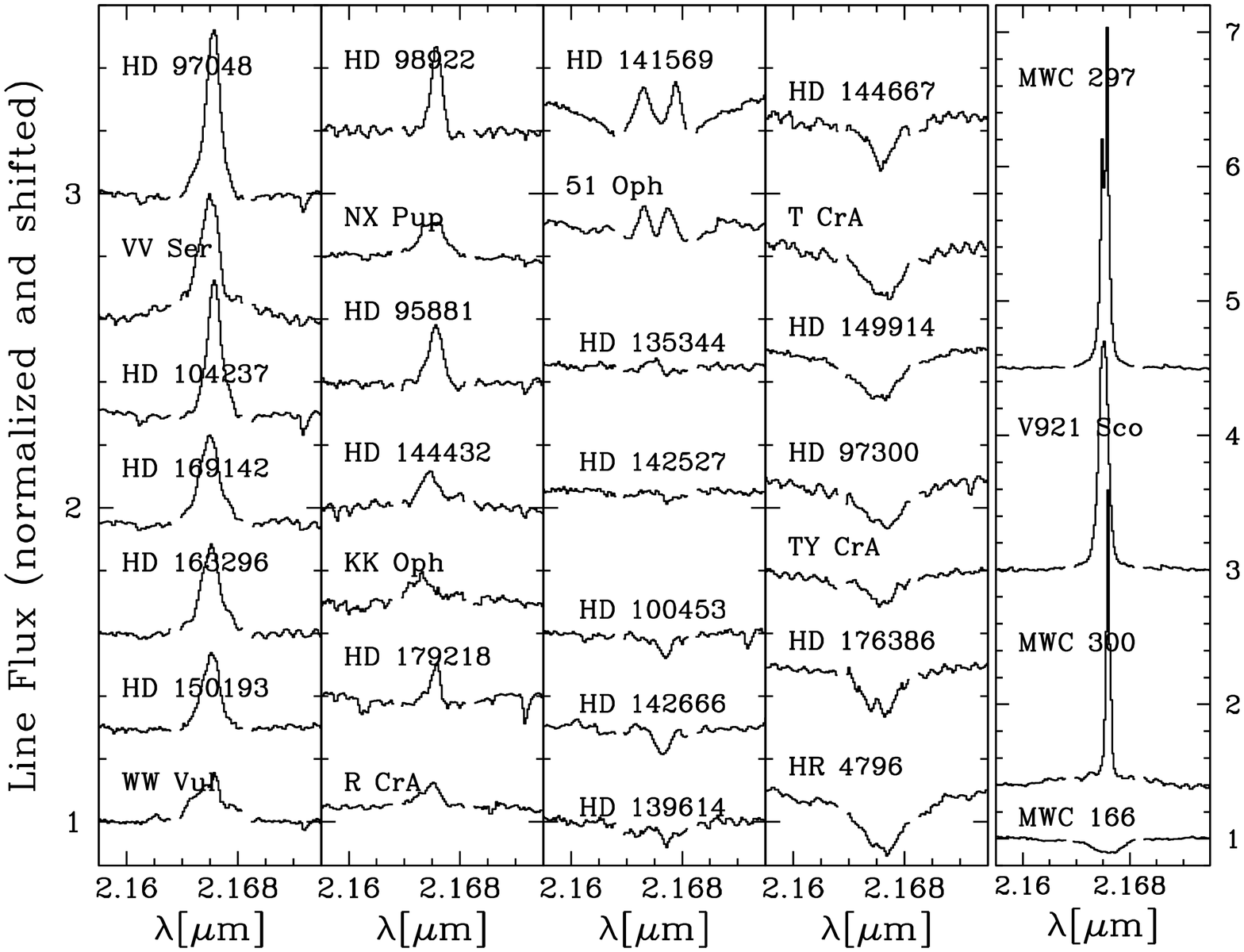,width=18cm,angle=-90} }
\caption {Observed \Brg\ profiles. The observed fluxes have been
normalized to the continuum and shifted for easier display. The stars
are roughly order according to the observed \Wobs, from strong
emission (left-top) to strong absorption (right-bottom). The
strongest peak emission (HD 97048) is about 44\% of the continuum;
the deepest absorption (HR 4796) is 15\% of the continuum.
The four early B stars are displayed in the fifth panel to the
right, with a different vertical scale.}
\label{fig_spectra}
\end{figure*}

The observed spectra, normalized to the continuum
level and shifted for an easier display,
are shown in Fig.~\ref{fig_spectra}.
They are ordered roughly according to the observed \Wobs, from strong
emission to strong absorption. In HAe, the
strongest peak emission (HD 97048) is about 44\% of the continuum;
the deepest absorption (HR 4796) is 15\% of the continuum.

The resolution of the spectra ($\sim$ 30 km s$^{-1}$) allows us to resolve
the lines in all objects. 
The observed line shape results from the
superposition of the circumstellar emission, the photospheric
absorption and the veiling K-band continuum, very likely from the
disk itself. One can see the signature of the
broad  photospheric absorption component in several objects, where
the K-band veiling is moderate or absent (e.g., HD 149914, HD 97300,
HR 4796). In some objects both the circumstellar emission 
and the photospheric absorption are clearly visible (e.g.,
HD 141569 and 51 Oph); in both objects the emission has a double-peaked
profile with peak-to-peak separation $\sim 300$ km s$^{-1}$.

In general, the emission lines are very broad, with
half width at 10\% of peak intensity (10\%HW) of 200--300 km s$^{-1}$.
VV Ser has very broad wings, which extends to $\pm$ 450 km s$^{-1}$ from the
line center. As a reference, the escape velocity of a typical
A0 star (\Mstar=2.3 \Msun, R=4 \Rsun)
is 330 km s$^{-1}$; the 10\%HW  of \Brg\ is often close to
this value.

In comparison, the three early B objects with emission have very strong 
lines. They are
relatively narrow  (10\%HW $\simless 120$ km s$^{-1}$)
when compared to the estimated escape velocity, which are
of  $\sim$ 200 km s$^{-1}$.

Clearly, the study of the profiles of the near-IR hydrogen emission lines in
these intermediate-mass objects  promises very interesting results, and
should be pursued further, possibly with even higher resolution spectrometers.

\end{document}